\documentclass[aps,prb,twocolumn,superscriptaddress,floatfix,longbibliography]{revtex4-2}

\usepackage{graphicx}
\usepackage{amsmath,amssymb}
\usepackage{bm}
\usepackage{xcolor}
\usepackage[colorlinks=true,allcolors=blue]{hyperref}

\begin{document}

\title{Conformal defects in a quasi-one-dimensional Kitaev chain:\\
an exactly parametrized Ising defect line and a relevant vacancy}

\author{R.~L.~Silva}
\affiliation{Departamento de Ci\^{e}ncias Naturais, Universidade Federal do Esp\'{i}rito Santo, S\~{a}o Mateus, ES 29932-540, Brazil}
\author{R.~C.~Silva}
\affiliation{Departamento de Ci\^{e}ncias Naturais, Universidade Federal do Esp\'{i}rito Santo, S\~{a}o Mateus, ES 29932-540, Brazil}
\author{A.~R.~Pereira}
\affiliation{Departamento de F\'isica, Universidade Federal de Vi\c{c}osa, Vi\c{c}osa, 36570-000 Minas Gerais, Brazil}

\date{\today}

\begin{abstract}
The quasi-one-dimensional Kitaev chain of coupled hexagonal plaquettes recently
introduced by Majumder is exactly solvable in terms of free Majorana fermions and
undergoes a topological transition at the exchange anisotropy
$J_x=\sqrt{2}\,J$. By eliminating the two-coordinated sites of each plaquette
we reduce the zero-energy problem to a uniform Majorana chain, which yields the
critical point, a single Majorana cone at $k=0$ (hence an Ising critical point
with central charge $c=\tfrac12$), and closed forms for every defect quantity
that follows. A weakened inter-plaquette bond $\lambda J_x$ is an exactly
marginal defect with transmission $T=4\lambda^2/(1+\lambda^2)^2$ at low
energy, that is, $T=\sin^2 2\phi$ with $\tan\phi=\lambda$: the bond traverses
the ferromagnetic defect line of the Ising conformal field theory from the
transparent point to the cut chain, and we give its exact energy dependence on
the lattice. The intra-plaquette bonds land on the same line with different
parametrizations: a $y$ or $z$ bond has $\tan\phi=(1+\lambda)/2$ and covers
only $T\in[0.64,1]$, because the plaquette provides a parallel path, whereas an
$x$ bond has $\tan\phi=(1+\lambda)/2\lambda$ and covers the strong-link half of
the line, $\phi\in(\pi/4,\pi/2)$. The identification is closed by an
independent observable: the coefficient of $\ln L$ in the entanglement entropy
across the defect follows the conformal-field-theory function
$\sigma(\sqrt T)/2$ of Brehm and Brunner to within $0.4\%$ along the whole
line, with no adjustable parameter. A vacancy behaves in the opposite way. It
never binds a plaquette flux, because plaquettes share no sites, but it leaves
its singly bonded neighbor as a conserved classical spin whose dangling
Majorana is a zero-energy resonance: the transmission vanishes as $T=E^2/2$ for
a vacancy on any site, so the vacancy is a relevant perturbation that cuts the
chain in the infrared, and it binds a Majorana zero mode only in the
topological phase. Finally, we show that
the clean ground state is exactly $4^{N_p}$-fold degenerate in the physical
spin Hilbert space, the degeneracy being the fixed-parity sector of the
$4N_p+2$ dangling Majoranas of the two-coordinated sites, and we identify the
local operators that act within this manifold. The degeneracy does not affect
the defect results, which live in the itinerant sector, but it must be
resolved before any density-matrix renormalization-group calculation on this
chain is meaningful.
\end{abstract}

\maketitle

\section{Introduction}

Defects in critical one-dimensional systems are classified by their fate under
the renormalization group. An irrelevant defect heals, a relevant one cuts the
system in two, and an exactly marginal one sits on a line of fixed points along
which a universal quantity, the transmission coefficient, varies continuously
\cite{KaneFisher1992a,KaneFisher1992b,EggertAffleck1992,Bachas2002,Quella2007}. For the Ising
universality class the marginal line is the ferromagnetic defect line, known on
the lattice since the work of Bariev and of McCoy and Perk on the
two-dimensional Ising model with a modified row of bonds
\cite{Bariev1979,McCoyPerk1980,Henkel1989,Delfino1994}, and whose boundary
state in the folded theory was constructed by Oshikawa and Affleck
\cite{OshikawaAffleck1997}. Along this line the entanglement entropy across
the defect carries a transmission-dependent prefactor, computed for
free-fermion chains by Eisler and Peschel \cite{EislerPeschel2010,Peschel2012}
and from the conformal field theory by Brehm and Brunner
\cite{BrehmBrunner2015} (see also Ref.~\cite{Sakai2008}). Lattice realizations
of the line are usually built on the transverse-field Ising chain itself, where
a modified bond is the natural defect \cite{Peschel2005,Igloi2009}. When one
asks instead what a \emph{vacancy}, or a modified bond in a \emph{different}
microscopic realization of the same conformal field theory, does, the answer
depends on the lattice, and the question of which lattice objects land on
which conformal defect is a legitimate one.

Kitaev-type spin models provide realizations of the Ising conformal field
theory that are microscopically far from the Ising chain: their critical
points are Majorana cones of an itinerant fermion moving in a static
$\mathbb Z_2$ gauge field \cite{Kitaev2006}. One-dimensional and ladder
versions of the Kitaev model have been studied for some time
\cite{Feng2007,Saket2010,Saket2013,DeGottardi2011}, and a new
quasi-one-dimensional member of this family was introduced recently by
Majumder \cite{Majumder2026}: a chain of hexagonal plaquettes cut out of the
honeycomb lattice and linked by one $x$ bond per plaquette. The model keeps
the bond-dependent exchange, the conserved plaquette operators, and the exact
Majorana representation of the honeycomb model, and it has a chiral (class
BDI \cite{Schnyder2008}) topological transition at the exchange anisotropy
$J_x=\sqrt{2J_yJ_z}$ separating a trivial phase from one with Majorana end
modes \cite{Majumder2026}. Because the plaquettes share no sites, the chain is
structurally a cactus graph, and the objects that are natural defects in the
honeycomb model---a vacancy, which there binds a flux
\cite{Willans2010,Willans2011,Santhosh2012}, or a modified bond---take on a
different character here.

In this paper we use this chain as a lattice laboratory for conformal defects,
with everything computed exactly from free Majorana fermions, checked against
exact diagonalization of the spin model, and, for the defect quantities,
derived in closed form. The key to the closed forms is that the
two-coordinated sites of a plaquette can be eliminated exactly, which reduces
the zero-energy problem to a uniform Majorana chain (Appendix~\ref{app:red}).
Our results are the following.
(i) The transition at $J_x=\sqrt{2J_yJ_z}$ is a \emph{single} Majorana cone
at $k=0$ with linear dispersion, so it is an Ising critical point with
$c=\tfrac12$; we confirm $c$ independently from the entanglement entropy.
(ii) A weakened inter-plaquette bond is an exactly marginal defect with
$T=4\lambda^2/(1+\lambda^2)^2$ at low energy, which is the Ising ferromagnetic
defect line with $\tan\phi=\lambda$, traversed from end to end; we give the
exact energy dependence of $T$ on the lattice. The intra-plaquette bonds land
on the same line with different parametrizations: a $y$ or $z$ bond has
$\tan\phi=(1+\lambda)/2$ and covers only $T\ge0.64$, whereas an $x$ bond has
$\tan\phi=(1+\lambda)/2\lambda$ and covers the strong-link half of the line.
The entanglement entropy across the defect follows the conformal-field-theory
prefactor $\sigma(\sqrt T)/2$ along the whole line to within $0.4\%$, with no
fitted parameter.
(iii) A vacancy is a relevant defect: its transmission vanishes as $T=E^2/2$
for every site, through a zero-energy Majorana resonance on the neighbor that
is left with a single bond, and the chain is cut in the infrared. There is no
flux binding, by structure. In the topological phase the vacancy binds a
Majorana zero mode.
(iv) The clean ground state is exactly $4^{N_p}$-fold degenerate in the
physical Hilbert space. The degeneracy comes from the $4N_p+2$ dangling
Majoranas of the two-coordinated sites, constrained to one fermion-parity
sector, and we give the local operators that act within it. This is the
lattice-level statement behind the $2^{2N}$ degeneracy noted in the extended
Majorana space in Ref.~\cite{Majumder2026}, and it has a practical consequence
for any numerical treatment of the chain, which we spell out.

The paper is organized as follows. Section~\ref{sec:model} presents the model,
its Majorana solution, and the identification of the critical point.
Section~\ref{sec:bond} treats the weakened bond and the entanglement test,
Sec.~\ref{sec:vacancy} the vacancy, and Sec.~\ref{sec:manifold} the degenerate
ground-state manifold. Section~\ref{sec:summary} summarizes. The exact
reduction of the plaquette and the derivations of the closed forms are
collected in Appendix~\ref{app:red}.

\section{Model and Majorana solution}
\label{sec:model}

\begin{figure*}[tb]
\includegraphics[width=\textwidth]{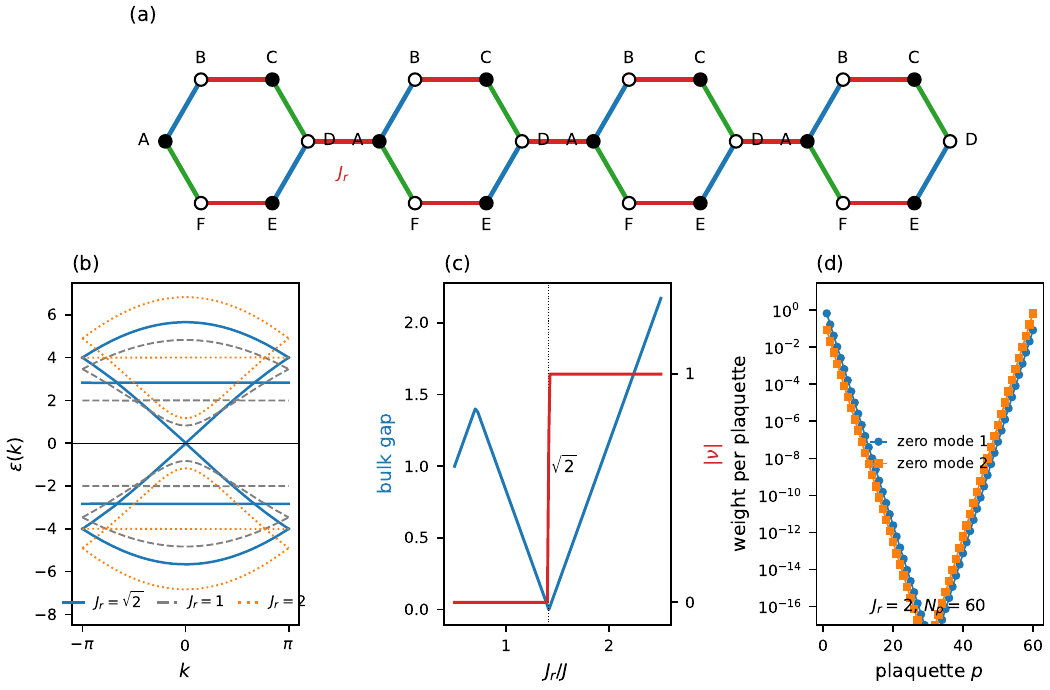}
\caption{(a) The chain: hexagonal plaquettes with sites $A,\dots,F$ and
bond-dependent couplings $J_x$ (red), $J_y$ (blue), and $J_z$ (green);
consecutive plaquettes are linked by a single $x$ bond $D(p)$--$A(p+1)$, also
of strength $J_x$. Filled and open circles mark the two sublattices. Sites
$B$, $C$, $E$, and $F$ have only two bonds. (b) Majorana bands
$\varepsilon(k)$ for $J_r\equiv J_x/J=1$, $\sqrt2$, and $2$ ($J_y=J_z=J=1$):
two flat bands at $\pm2J_x$ and, at $J_r=\sqrt2$, a single cone at $k=0$.
(c) Bulk gap and winding number $|\nu|$ versus $J_r$. (d) Weight per plaquette
of the two zero modes of an open chain of $N_p=60$ plaquettes at $J_r=2$.}
\label{fig:modelo}
\end{figure*}

The model \cite{Majumder2026} is a chain of $N_p$ hexagonal plaquettes with
sites $A,B,C,D,E,F$ [Fig.~\ref{fig:modelo}(a)],
\begin{align}
H=\sum_p\big[&J_y\sigma^y_{A}\sigma^y_{B}+J_x\sigma^x_{B}\sigma^x_{C}
+J_z\sigma^z_{C}\sigma^z_{D}+J_y\sigma^y_{D}\sigma^y_{E}\nonumber\\
&+J_x\sigma^x_{E}\sigma^x_{F}+J_z\sigma^z_{F}\sigma^z_{A}\big]_p
+J_x\sum_p\sigma^x_{D(p)}\sigma^x_{A(p+1)},
\label{eq:H}
\end{align}
where $\sigma^a$ are Pauli matrices. The inter-plaquette bond is an $x$ bond
of the honeycomb lattice and carries the same coupling $J_x$ as the two
intra-plaquette $x$ bonds. Following Ref.~\cite{Majumder2026} we set
$J_y=J_z=J=1$ and use the anisotropy $J_r\equiv J_x/J$ as the control
parameter; a weakened bond will be written as $\lambda$ times its clean
value. Each site has at most one bond of each type, so the Kitaev
construction applies \cite{Kitaev2006}: $\sigma^a_j=i b^a_j c_j$ with four
Majoranas per site and the constraint $D_j=b^x_jb^y_jb^z_jc_j=1$; the bond
operators $u_{jk}=ib^a_jb^a_k$ are conserved and
\begin{equation}
H=\frac{i}{4}\sum_{jk}A_{jk}c_jc_k,\qquad A_{jk}=2K_{jk}u_{jk},
\label{eq:majorana}
\end{equation}
with $K_{jk}$ the coupling of the bond $jk$. The single-particle energies are
the positive eigenvalues of the Hermitian matrix $iA$, whose hopping
amplitudes are $t_a=2J_a$ on a bond of type $a$. The lattice is bipartite,
with sublattices $\{A,C,E\}$ and $\{B,D,F\}$, and the inter-plaquette bond
joins the two sublattices, so the gauge $u_{jk}=+1$ from sublattice 1 to 2 is
consistent and defines the flux-free sector. The plaquette operators
$W_p=\sigma^x_A\sigma^z_B\sigma^y_C\sigma^x_D\sigma^z_E\sigma^y_F$ commute with
$H$; the inter-plaquette bonds close no loop, so there is no other conserved
flux, and the ground state has $W_p=+1$ on every plaquette
\cite{Majumder2026}. We verified all of this against exact diagonalization of
the spin Hamiltonian: the free-Majorana ground-state energy agrees with that
of the spin model to nine digits for $N_p=1$ and $2$ and for three sets of
couplings, and $\langle W_p\rangle=+1$ in the spin ground state. These checks
fix the numerical factor in Eq.~(\ref{eq:majorana}) and the gauge, and
everything below rests on them.

The single-particle problem is the $6\times6$ Bloch matrix
$h(k)=i\sum_d A_d e^{ikd}$. Its bands are shown in Fig.~\ref{fig:modelo}(b).
Two bands are flat at $\varepsilon=\pm t_x=\pm2J_x$ for every $J_r$; in the
units of Ref.~\cite{Majumder2026}, where the exchange is written with
spin-$\tfrac12$ operators, these are the flat bands at $\pm J_x$. The
remaining four bands disperse, and the gap closes only at $J_r=\sqrt2$.

The structure of the spectrum has a simple origin, which we exploit
throughout the paper (Appendix~\ref{app:red}). The sites $B,C,E,F$ have two
bonds each and can be eliminated exactly at any energy $E$. Each arm of the
hexagon, $A$--$B$--$C$--$D$ or $A$--$F$--$E$--$D$, then becomes an effective
coupling between $A$ and $D$ of magnitude $t_yt_xt_z/|E^2-t_x^2|$, together
with energy-dependent on-site terms; the two arms carry the same amplitude in
the gauge $u=+1$ and add. At $E=0$ the on-site terms vanish and the plaquette
collapses to a single bond of strength $2t_yt_z/t_x$ between $A$ and $D$, so
the zero-energy problem is a two-site chain with alternating hoppings
$2t_yt_z/t_x$ (through the plaquette) and $t_x$ (between plaquettes). This
chain is gapless if and only if the two hoppings are equal,
\begin{equation}
t_x^2=2t_yt_z\quad\Longleftrightarrow\quad J_x=\sqrt{2J_yJ_z},
\label{eq:critical}
\end{equation}
which is the critical anisotropy found numerically in Ref.~\cite{Majumder2026}
and equals $J_r=\sqrt2$ for $J_y=J_z=J$. For $J_x<\sqrt{2J_yJ_z}$ the
intra-plaquette hopping dominates and the open chain is trivial; for
$J_x>\sqrt{2J_yJ_z}$ the inter-plaquette hopping dominates, the end sites
$A(1)$ and $D(N_p)$ are weakly attached, and the chain is topological with
Majorana end modes, exactly as in the dimerized Kitaev wire \cite{Kitaev2001}.
The flat bands at $\pm t_x$ are the energies at which the elimination is
singular: they are the antisymmetric combinations of the inner sites of the
two arms, which decouple from $A$ and $D$. At the critical point the
zero-energy chain is uniform with two sites per unit cell, so it has exactly
one gap closing in the reduced Brillouin zone, at $k=0$ in our convention: a
\emph{single} Majorana cone. Its velocity is reduced from the bare hopping
$t_x$ by the energy dependence of the eliminated sites,
$v=t_x^3/(t_x^2+t_y^2+t_z^2)=J_x$ for $J_y=J_z$ [Eq.~(\ref{eq:v})], that is,
$v=\sqrt2\,J$. A scan of $4\times10^4$ points across the Brillouin zone
confirms all of this: at $J_r=\sqrt2$ the gap closes at exactly one momentum,
$k=0$, and the dispersion there is linear, $\varepsilon=\pm v|k|$ with
$v/J=1.414214$ constant over three decades of $k$. Since $k=0$ is its own
time-reversal partner, the critical point is one free Majorana fermion, which
is the Ising conformal field theory with $c=\tfrac12$. This is the fact on
which the rest of the paper depends.

The chiral symmetry is the sublattice bipartition; in the sublattice basis the
diagonal blocks of $h(k)$ vanish, and the winding number of $\det q(k)$ of the
off-diagonal block is $\nu=0$ for $J_r<\sqrt2$ and $|\nu|=1$ for $J_r>\sqrt2$
[Fig.~\ref{fig:modelo}(c)], as found in Ref.~\cite{Majumder2026}. An open
chain in the topological phase has two Majorana modes at
$|\varepsilon|<10^{-15}$ with $98\%$ of their weight in the three outermost
plaquettes [Fig.~\ref{fig:modelo}(d)]. Close to the transition their
localization length grows, and at $J_r=1.5$ they are still hybridized
($\varepsilon=3\times10^{-4}$) in a chain of 360 sites.

\section{A weakened bond realizes the Ising defect line}
\label{sec:bond}

\begin{figure}[tb]
\includegraphics[width=\columnwidth]{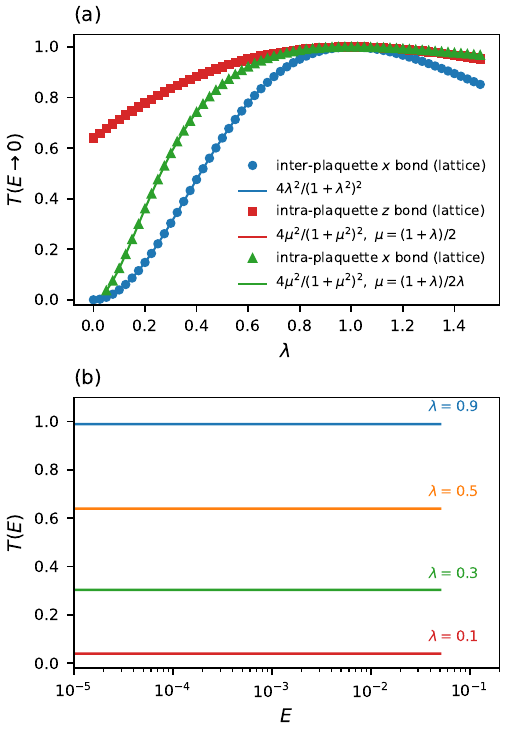}
\caption{Transmission through a weakened bond at the critical point
$J_r=\sqrt2$. (a) $T(E\to0)$ versus the bond factor $\lambda$ for the
inter-plaquette $x$ bond, the intra-plaquette $z$ bond ($C$--$D$), and the
intra-plaquette $x$ bond ($B$--$C$); symbols are the lattice Fisher--Lee
result at $E=10^{-4}$, lines are the closed forms of Eqs.~(\ref{eq:Tinter}),
(\ref{eq:Tintra}), and (\ref{eq:Tintrax}). (b) $T(E)$ for the inter-plaquette
bond: constant over four decades, the signature of an exactly marginal
defect; the exact lattice correction, Eq.~(\ref{eq:TinterE}), is below
$10^{-6}$ in this window.}
\label{fig:transmissao}
\end{figure}

We now place a single defect in the critical chain and compute the
transmission of a Majorana fermion through it, using the Fisher--Lee formula
\cite{FisherLee1981} with the surface Green's functions of the two
semi-infinite leads obtained by the decimation scheme of L\'opez Sancho
\emph{et al.} \cite{Sancho1985}. Without a defect, $T=1.000000$ from
$E=10^{-2}$ down to $10^{-5}$: there is a single channel, the cone. The
reduction of Appendix~\ref{app:red} makes every result of this section
analytic. At the critical point the surface Green's function of the
semi-infinite chain at its terminal site is exactly that of a uniform chain
with hopping $t_x$, $g(E)=[E-i\sqrt{4t_x^2-E^2}]/2t_x^2$
[Eq.~(\ref{eq:gsurf})], so a modified link between two half-chains is a
textbook single-link scattering problem, and a modified bond inside a
plaquette is the same problem with an effective link strength.

We first multiply the inter-plaquette bond between two given plaquettes by
$\lambda$. The exact lattice transmission is [Appendix~\ref{app:red},
Eq.~(\ref{eq:TinterE_app})]
\begin{equation}
T_{\rm inter}(\lambda,E)=\frac{4\lambda^2 s^2}{(1-\lambda^2)^2+4\lambda^2s^2},
\qquad s^2=1-\frac{E^2}{4t_x^2},
\label{eq:TinterE}
\end{equation}
which agrees with the numerical Fisher--Lee result to $10^{-9}$, the accuracy
set by the broadening, for all $\lambda$ and all $|E|<2t_x$. Its low-energy
limit,
\begin{equation}
T_{\rm inter}(\lambda)=\frac{4\lambda^2}{(1+\lambda^2)^2},
\label{eq:Tinter}
\end{equation}
is approached with corrections of relative order $E^2/4t_x^2=E^2/32$, below
$10^{-6}$ for $E\le10^{-2}$ [Fig.~\ref{fig:transmissao}(b)]: the defect
neither heals nor cuts; it is exactly marginal, and $T(\lambda)$ labels a line
of fixed points. Equation~(\ref{eq:Tinter}) covers the whole interval from
$T=1$ at $\lambda=1$ to $T=0$ at $\lambda=0$, the cut chain
[Fig.~\ref{fig:transmissao}(a)]. It is the same function of the bond factor
as the transmission through a single modified hopping in the critical
transverse-field Ising or XX chain \cite{EislerPeschel2010,Peschel2012}: at
low energy, the inter-plaquette bond behaves exactly as a bond defect of the
Ising chain, even though the two lattices are very different away from the
cone. Here this is not a coincidence but a consequence of
Eq.~(\ref{eq:gsurf}).

We next weaken a bond inside a plaquette. Since at $E\to0$ the plaquette is a
single effective link whose strength is the sum of the two arms
(Appendix~\ref{app:red}), a bond factor $\lambda$ on one arm turns the
plaquette into a link of relative strength $\mu$ in an otherwise uniform
chain, and $T=4\mu^2/(1+\mu^2)^2$. For the $z$ bond $C$--$D$ (or, by the
mirror symmetry of the plaquette, the $z$ bond $F$--$A$, and likewise for the
two $y$ bonds) the arm strength is linear in $\lambda$ and
\begin{equation}
T_{\rm intra}^{(y,z)}(\lambda)=\frac{4\mu^2}{(1+\mu^2)^2},\qquad
\mu=\frac{1+\lambda}{2}.
\label{eq:Tintra}
\end{equation}
Removing the bond altogether ($\lambda=0$) leaves $T=0.64$ rather than $0$:
the other arm of the hexagon offers a parallel path, and the effective bond
factor $\mu$ is the arithmetic mean of the factors of the two arms, $\lambda$
and $1$. A $y$ or $z$ intra-plaquette bond therefore covers only the segment
$T\in[0.64,1]$ of the line. The intra-plaquette $x$ bonds ($B$--$C$ and
$E$--$F$) are different: the $x$ bond sits in the middle of its arm, and the
arm strength $t_yt_z/(\lambda t_x)$ \emph{grows} as the bond is weakened, so
\begin{equation}
T_{\rm intra}^{(x)}(\lambda)=\frac{4\mu^2}{(1+\mu^2)^2},\qquad
\mu=\frac{1+\lambda}{2\lambda},
\label{eq:Tintrax}
\end{equation}
valid for $E\ll\lambda t_x$. Here $\mu\ge1$ for $\lambda\le1$, and $\mu\to\infty$
as $\lambda\to0$: weakening an intra-plaquette $x$ bond also cuts the chain,
but through an effective link that is \emph{stronger} than the bulk. All three
closed forms agree with the lattice values to the accuracy of the broadening
[Fig.~\ref{fig:transmissao}(a)].

All of these bonds realize the ferromagnetic defect line of the Ising
conformal field theory. For that line the transmission is $T=\sin^2 2\phi$,
where $\phi\in[0,\pi/2]$ parametrizes the boundary state of the folded theory
\cite{OshikawaAffleck1997,BrehmBrunner2015}, with $\phi=\pi/4$ the transparent
point and $\phi=0,\pi/2$ the two cut chains. Equation~(\ref{eq:Tinter}) is
$\sin^22\phi$ with $\tan\phi=\lambda$: the inter-plaquette bond runs over
$\phi\in[0,\pi/4]$. The intra-plaquette $y$ and $z$ bonds have
$\tan\phi=(1+\lambda)/2$ and run over $\phi\in[\arctan\tfrac12,\pi/4]$, and
the intra-plaquette $x$ bonds have $\tan\phi=(1+\lambda)/2\lambda$ and run
over the complementary half $\phi\in(\pi/4,\pi/2)$. The three bond types of
the lattice thus cover the whole line: the inter-plaquette bond alone covers the
weak-link half $\phi\in[0,\pi/4]$, the intra-plaquette $y$ and $z$ bonds retrace
the upper part of that half, and the intra-plaquette $x$ bonds are the only ones
that reach the strong-link half $\phi\in(\pi/4,\pi/2)$. On this line the boundary entropy $g$ \cite{AffleckLudwig1991}
is constant: the identity component of the boundary state has coefficient
one for every $\phi$ [Eq.~(3.36) of Ref.~\cite{OshikawaAffleck1997}],
consistent with an exactly marginal deformation, which cannot change $g$. The
universal quantity that does vary with the transmission is the prefactor of
the entanglement entropy across the defect.

\begin{figure}[tb]
\includegraphics[width=\columnwidth]{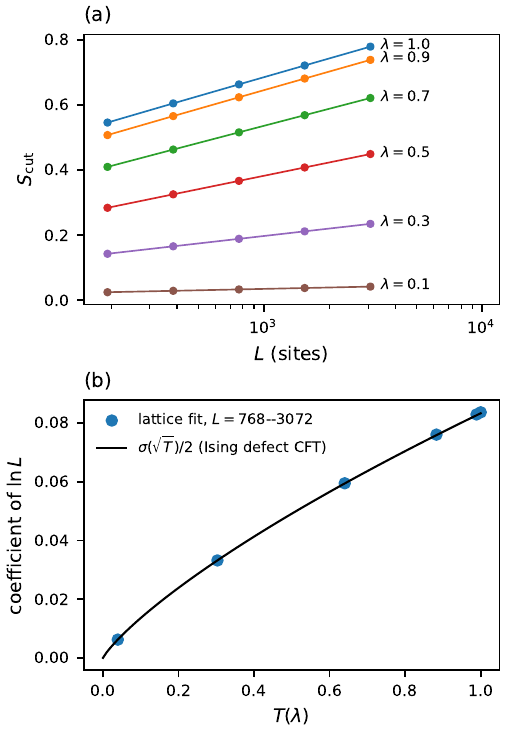}
\caption{Entanglement entropy across the inter-plaquette defect in an open
critical chain of $L=6N_p$ sites, bipartitioned exactly at the defect.
(a) $S_{\rm cut}$ versus $L$ for several $\lambda$. (b) Coefficient of
$\ln L$, fitted in $L=768$--$3072$, versus the transmission $T(\lambda)$ of
Eq.~(\ref{eq:Tinter}), compared with the conformal-field-theory prediction
$\sigma(\sqrt T)/2$ of Ref.~\cite{BrehmBrunner2015} [Eq.~(\ref{eq:sigma})]. No
parameter is adjusted.}
\label{fig:entropia}
\end{figure}

That prefactor provides an independent test of the identification. For the
Ising defect line, the entanglement entropy of the two halves of an infinite
chain joined by the defect is $S=\sigma(s)\ln L+{\rm const}$ with $s=\sqrt T$
and \cite{BrehmBrunner2015}
\begin{equation}
\sigma(s)=\frac{2}{\pi^2}\int_0^\infty du\,u\Big(\sqrt{1+\frac{s^2}{\sinh^2u}}-1\Big),
\label{eq:sigma}
\end{equation}
which reduces to $\sigma(1)=\tfrac16=c/3$ without a defect
\cite{Holzhey1994,Vidal2003,CalabreseCardy2004} and vanishes as $-T\ln T$
when the chain is cut; it coincides with one half of the effective central
charge obtained by Eisler and Peschel for the XX chain
\cite{EislerPeschel2010,Peschel2012}. For an open chain cut at the defect the
coefficient is halved. We computed the entanglement entropy of the
free-Majorana ground state of open chains of $N_p=32$ to $512$ plaquettes,
bipartitioned at the defect, from the correlation matrix \cite{Peschel2003},
and fitted the coefficient of $\ln L$ in $N_p=128$--$512$. This is the entropy
of the itinerant Majorana sector in the fixed gauge; the dangling modes
discussed in Sec.~\ref{sec:manifold} contribute only size-independent
constants and do not affect the coefficient of $\ln L$.
Figure~\ref{fig:entropia} compares the fitted coefficient with
$\sigma(\sqrt T)/2$, with $T$ taken from Eq.~(\ref{eq:Tinter}). The agreement
is between $0.1\%$ and $0.4\%$ along the entire line (Table~\ref{tab:sigma}),
the residual being of the size expected from $1/L$ corrections in a
three-size fit. At $\lambda=1$ the coefficient gives $12a=1.004$, that is,
$c=\tfrac12$ again, by a route independent of the band structure. Three
separate calculations---the scattering matrix, the entropy of the correlation
matrix, and the conformal-field-theory integral---are tied together by the
single lattice parameter $\lambda$ with nothing adjusted. This closes the
chain of identifications: single cone $\to$ Ising $\to$ ferromagnetic defect
line $\to$ $\tan\phi=\lambda$.

\begin{table}[tb]
\caption{Coefficient $a$ of $\ln L$ in the entanglement entropy across the
inter-plaquette defect (lattice, $N_p=128$--$512$) compared with
$\sigma(\sqrt T)/2$ from Eq.~(\ref{eq:sigma}), with $T$ from
Eq.~(\ref{eq:Tinter}).}
\label{tab:sigma}
\begin{ruledtabular}
\begin{tabular}{cccc}
$\lambda$ & $T$ & $a$ (lattice) & $\sigma(\sqrt T)/2$ \\
\hline
1.0 & 1.0000 & 0.08366 & 0.08333 \\
0.9 & 0.9890 & 0.08293 & 0.08264 \\
0.7 & 0.8828 & 0.07605 & 0.07588 \\
0.5 & 0.6400 & 0.05948 & 0.05942 \\
0.3 & 0.3030 & 0.03319 & 0.03321 \\
0.1 & 0.0392 & 0.00622 & 0.00623 \\
\end{tabular}
\end{ruledtabular}
\end{table}

\section{A vacancy is a relevant defect}
\label{sec:vacancy}

\begin{figure}[tb]
\includegraphics[width=\columnwidth]{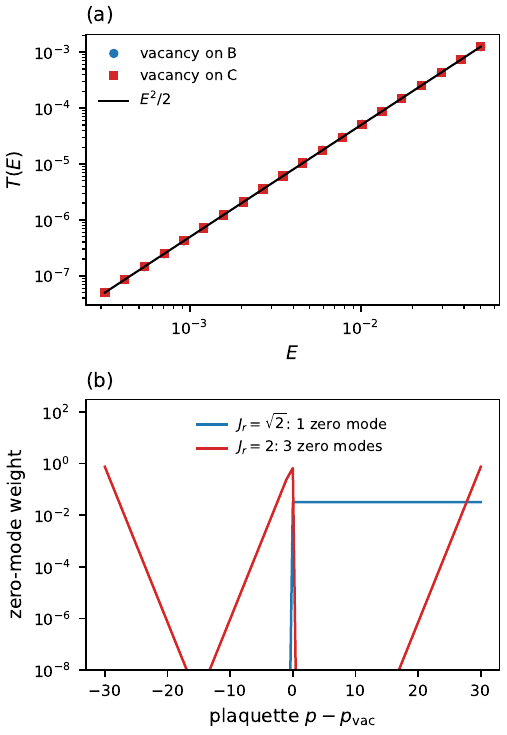}
\caption{A single vacancy. (a) Transmission at the critical point through a
vacancy on site $B$ or $C$ (the two inequivalent sites that keep the chain
connected), compared with $E^2/2$. (b) Weight per plaquette of the zero modes
of an open chain of $N_p=61$ plaquettes with the vacancy on site $C$ of the
central plaquette, at the critical point and in the topological phase.}
\label{fig:vacancia}
\end{figure}

Removing a site removes four Majoranas and, in the free-fermion language, one
row and one column of $A$. The structure of the chain settles several things
before any number is computed. Removing $A$ or $D$ removes the only
inter-plaquette bond and cuts the chain literally, $T\equiv0$. Removing $B$,
$C$, $E$, or $F$ opens the hexagon into a path and the chain stays connected;
by the symmetry $A\leftrightarrow D$, $B\leftrightarrow E$,
$C\leftrightarrow F$ there are two inequivalent connected vacancies, $B$ and
$C$. Unlike the honeycomb model, where a vacancy binds a flux by merging the
three plaquettes around it \cite{Willans2010,Willans2011}, here a vacancy
cannot bind a flux: plaquettes share no sites, so the vacancy destroys its own
plaquette instead of merging neighbors. What is left is a neighbor with a
\emph{single} bond ($C$ for a vacancy on $B$, $B$ for a vacancy on $C$), whose
remaining Pauli component commutes exactly with $H$ (we checked
$[H,\sigma^z_C]=0$ in the spin model) and acts as a conserved classical spin,
a longitudinal field of either sign on its one neighbor. In Majorana terms it
is an itinerant Majorana $c$ hanging from the chain by a single bond. The
free-Majorana energy with the vacancy, in the gauge $u=+1$, agrees with the
spin exact diagonalization to $10^{-9}$ both for a vacancy at the end
($N_p=2$, 11 spins) and in the interior ($N_p=3$, 17 spins, Lanczos), so the
description remains exact.

At the critical point the vacancy is not a conformal defect: the transmission
vanishes at low energy for every site, including the ones that keep the chain
connected, with the law
\begin{equation}
T(E)=E^2/2
\label{eq:Tvac}
\end{equation}
(fitted exponent $2.0000$, amplitude $0.4998$ from $E=3\times10^{-2}$ to
$3\times10^{-4}$; Fig.~\ref{fig:vacancia}(a)). The mechanism is the dangling
Majorana: a single Majorana coupled to the chain by one bond is a resonance at
zero energy, and at $E=0$ it reflects the incoming cone mode perfectly, as a
Fano antiresonance \cite{Fano1961}. The law follows from the reduction of
Appendix~\ref{app:red}. The dangling site contributes a self-energy $t_d^2/E$
to its neighbor, with $t_d=t_y$ or $t_z$ the bond that survives; the
remaining arm of the damaged plaquette is a link of strength $t_yt_z/t_x$; and
at low energy the transmission through this link next to a site with a
divergent on-site term is $T=(2E\,t_yt_z/t_xt_d^2)^2$
[Eq.~(\ref{eq:Tvac_app})], which at the critical point with $t_y=t_z=t_d$
reduces to $T=4E^2/t_x^2=E^2/2$. The same law holds for a vacancy on $B$, $C$, $E$,
or $F$, since the four cases differ only by the mirror symmetries of the
plaquette. In the language of the conformal field
theory, a boundary Majorana $\gamma$ coupled to the bulk fermion $\psi(0)$ is
an operator of dimension $\tfrac12$, hence relevant, and flows to the cut
fixed point. The two descriptions agree: a weakened bond is marginal and sits
on a line, whereas a vacancy is relevant and cuts the chain in the infrared;
they are different objects from the conformal point of view even though both
are single-site defects on the lattice.

In the topological phase the vacancy does bind a Majorana zero mode
[Fig.~\ref{fig:vacancia}(b)]. At $J_r=2$ an open chain of $61$ plaquettes with
the vacancy on site $C$ of the central plaquette has three modes at
$|\varepsilon|<10^{-6}$: the two end modes, and one with $98\%$ of its weight
in the five plaquettes around the vacancy and $67\%$ on the single-bond site.
At the critical point there is one exact zero mode, forced by the odd number
$6N_p-1$ of Majoranas, but it is delocalized ($81\%$ in the bulk, $0.8\%$ on
the single-bond site): there is no bound state at criticality.

\section{The degenerate ground-state manifold}
\label{sec:manifold}

Exact diagonalization of the clean open chain gives a ground state that is
\emph{exactly} $4$-fold degenerate for $N_p=1$ and $16$-fold degenerate for
$N_p=2$, that is, $4^{N_p}$-fold, separated from the next level by a gap of
order $J$ ($0.54J$ and $0.49J$, respectively); the degeneracy is preserved (up
to the factor 2 of the classical spin) by a vacancy. It is not a flux
degeneracy: flipping a $W_p$ costs $0.3$--$0.7\,J$ in all sizes and changes
the single-particle spectrum. Its origin is the coordination. Sites $B,C,E,F$
have only two bonds, so one Pauli component on each of them ($\sigma^z_B$,
$\sigma^y_C$, $\sigma^z_E$, $\sigma^y_F$) never appears in $H$: the
corresponding $b$ Majorana is absent from Eq.~(\ref{eq:majorana}) and is an
exact zero mode of the extended problem, the same mechanism that produces the
local conserved quantities of the one-dimensional Kitaev models of
Refs.~\cite{Feng2007,DeGottardi2011}. This is the origin of the $2^{2N}$
degeneracy of the extended Majorana space noted in Ref.~\cite{Majumder2026};
the statement here is that the degeneracy survives the projection $D_j=1$ to
the physical spin space \cite{Pedrocchi2011,Zschocke2015}, where it is
$4^{N_p}$, and that the count is exact. On an open chain the end sites $A(1)$
and $D(N_p)$ also lack their inter-plaquette bond and contribute a dangling
$b^x$ each, for a total of $4N_p+2$ dangling Majoranas, that is, $2N_p+1$
fermion modes; the global constraint $\prod_jD_j=1$, which acts trivially on
the $u$'s, fixes the parity of these modes jointly with that of the itinerant
sector, and the physical ground manifold is the fixed-parity sector, of
dimension $2^{2N_p+1}/2=4^{N_p}$.

The operators that move within the manifold are the gauge-invariant bilinears
of two dangling Majoranas dressed by the bond string between them. They are
short products of Pauli matrices that commute exactly with $H$, and we found
the complete set by direct search: within each plaquette,
$\sigma^y_B\sigma^z_C$ and $\sigma^y_E\sigma^z_F$; through the
three-coordinated sites, $\sigma^x_C\sigma^x_D\sigma^x_E$ and
$\sigma^x_F\sigma^x_A\sigma^x_B$; across the inter-plaquette bond,
$\sigma^x_{C(p)}\sigma^x_{D(p)}\sigma^z_{A(p+1)}\sigma^x_{B(p+1)}$; and at the
two ends, $\sigma^z_A\sigma^x_B$, $\sigma^y_A\sigma^x_F$ and
$\sigma^x_C\sigma^y_D$, $\sigma^z_D\sigma^x_E$. Applied repeatedly to any one
ground state they generate the full $4^{N_p}$-dimensional manifold, which we
verified numerically up to $N_p=3$ ($64$ states, residual $10^{-11}$).

None of the defect results above depends on this manifold: the transmission
and the $\ln L$ coefficient of the entropy are properties of the itinerant
$c$ sector, and the dangling modes add only size-independent constants to the
entropy. The practical consequence is for numerics. A density-matrix
renormalization-group calculation on this chain converges to an arbitrary
vector of the manifold: we find that three random starts at $N_p=2$ reach the
same energy to $10^{-10}$ but three different states, with mutual overlaps of
$0.3$. Any dynamical or correlation quantity computed from ``the'' ground state
of Eq.~(\ref{eq:H}), such as the structure factors of
Ref.~\cite{Majumder2026}, depends on which vector of the manifold the
algorithm happened to reach, unless the degeneracy is lifted by a perturbation
or the observable is manifestly independent of the dangling sector. How a
uniform field lifts the manifold is a question in its own right, since the
first-order matrix elements vanish and the lifting occurs at second order
through the critical itinerant sector; we treat it in a separate paper.

\section{Summary and outlook}
\label{sec:summary}

The quasi-one-dimensional Kitaev chain realizes the Ising conformal field
theory through a single Majorana cone, and its bonds realize the Ising defect
line with analytic parametrizations that follow from a single exact step, the
elimination of the two-coordinated sites of each plaquette: $\tan\phi=\lambda$
for the inter-plaquette bond, which covers the line from transparency to the
cut; $\tan\phi=(1+\lambda)/2$ for an intra-plaquette $y$ or $z$ bond, which
covers only $T\ge0.64$ because the other arm of the plaquette provides a
parallel path; and $\tan\phi=(1+\lambda)/2\lambda$ for an intra-plaquette $x$
bond, which covers the strong-link half of the line. The same step gives the
critical anisotropy $J_x=\sqrt{2J_yJ_z}$, the velocity of the cone, and the
exact energy dependence of the inter-plaquette transmission on the lattice.
The conformal prefactor of the entanglement entropy follows the lattice
transmission along the whole line to within $0.4\%$ with nothing fitted. A
vacancy is the opposite kind of object: it binds no flux, it leaves a
conserved classical spin whose dangling Majorana is a zero-energy resonance,
and it cuts the chain in the infrared with $T=4E^2/t_x^2=E^2/2$, binding a
Majorana zero mode only in the topological phase. The clean ground state is
exactly $4^{N_p}$-fold degenerate in the physical Hilbert space, a fact that
leaves the defect physics untouched but that any numerical treatment of the
chain has to confront.

Two of these statements are, we believe, general beyond this chain. First,
the transmission of a weakened link between two segments of a Kitaev-type
chain whose critical point is a single cone is the Ising-chain law
$4\lambda^2/(1+\lambda^2)^2$, independently of the microscopic structure of
the segments; the lattice enters only through the effective link factor, as
the two intra-plaquette results illustrate. Second, in any Kitaev-type
lattice in which plaquettes share no sites, a vacancy cannot bind a flux and
instead leaves a dangling itinerant Majorana, so it is a relevant rather than
a marginal defect. Kitaev ladders and other cactus-like geometries
\cite{Feng2007,Saket2010,Saket2013,DeGottardi2011} offer further tests of both
statements.

\begin{acknowledgments}
A.~R.~P. thanks INCT/CNPq -- Spintr\^onica e Nanoestruturas Magn\'eticas
Avan\c{c}adas (INCT-SpinNanoMag) and Rede Mineira de Nanomagnetismo/FAPEMIG.
This work was supported by the Brazilian agencies CNPq, FAPEMIG, CAPES and
FAPES.
\end{acknowledgments}

\appendix

\section{Exact reduction of the plaquette and closed forms}
\label{app:red}

In the gauge $u=+1$ the matrix $iA$ of Eq.~(\ref{eq:majorana}) is a Hermitian
hopping matrix with amplitudes of magnitude $t_a=2J_a$ on the bonds of type
$a=x,y,z$ and zero on-site energies. We write $h$ for it and $E$ for the
energy; the resolvent is $G(E)=(E+i0^+-h)^{-1}$.

\paragraph*{Elimination of an arm.}
Consider a path $A$--$B$--$C$--$D$ with amplitudes $h_{AB}$, $h_{BC}$,
$h_{CD}$ and no other bonds on $B$ and $C$. Eliminating $B$ and $C$ at energy
$E$ (Gaussian elimination on the resolvent, or equivalently the Schur
complement) adds to the remaining sites the energy-dependent terms
\begin{align}
\Sigma_{AA}&=\frac{|h_{AB}|^2E}{E^2-|h_{BC}|^2},\qquad
\Sigma_{DD}=\frac{|h_{CD}|^2E}{E^2-|h_{BC}|^2},\nonumber\\
\Sigma_{AD}&=\frac{h_{AB}h_{BC}h_{CD}}{E^2-|h_{BC}|^2}.
\label{eq:elim}
\end{align}
A plaquette has two such arms between $A$ and $D$, $A$--$B$--$C$--$D$ with
bond types $(y,x,z)$ and $A$--$F$--$E$--$D$ with bond types $(z,x,y)$. Both
run from sublattice 1 to 2 to 1 to 2, so in the gauge $u=+1$ their amplitude
products $h_{AB}h_{BC}h_{CD}$ and $h_{AF}h_{FE}h_{ED}$ are equal, and the two
contributions to $\Sigma_{AD}$ add. With $J_y=J_z$ the plaquette therefore
reduces to the two sites $A$ and $D$ with
\begin{equation}
\epsilon(E)=\frac{(t_y^2+t_z^2)\,E}{E^2-t_x^2},\qquad
|t_{\rm eff}(E)|=\frac{2t_yt_xt_z}{|E^2-t_x^2|},
\label{eq:eff}
\end{equation}
on both sites and between them, respectively, while the inter-plaquette bond
$D(p)$--$A(p+1)$ keeps its amplitude $t_x$. The elimination is singular at
$E=\pm t_x$; these are the two flat bands of Fig.~\ref{fig:modelo}(b), the
combinations of the inner sites of the two arms that are antisymmetric under
the mirror exchanging the arms and therefore decouple from $A$ and $D$.

\paragraph*{Critical point, single cone, and velocity.}
At $E=0$ the on-site terms vanish and $|t_{\rm eff}(0)|=2t_yt_z/t_x$. The
zero-energy problem is a chain with two sites per cell and alternating
hoppings $2t_yt_z/t_x$ and $t_x$, which is gapless if and only if they are
equal, Eq.~(\ref{eq:critical}). Then the chain is uniform with two sites per
unit cell of length $1$ and the off-diagonal Bloch element is
$t_x(1-e^{ik})$ up to phases, whose modulus $2t_x|\sin(k/2)|$ vanishes once,
at $k=0$: a single cone. Away from $E=0$ the effective problem is
$\det[E-\epsilon(E)-h_{\rm eff}(k,E)]=0$; to first order in $E$,
$\epsilon(E)\simeq-(t_y^2+t_z^2)E/t_x^2$ and
$|t_{\rm eff}(E)|\simeq2t_yt_z/t_x$, so
$E\,[1+(t_y^2+t_z^2)/t_x^2]=\pm t_x|k|$ and
\begin{equation}
v=\frac{t_x^3}{t_x^2+t_y^2+t_z^2}=\frac{t_x}{2}=J_x
\quad(J_y=J_z,\ t_x^2=2t_yt_z),
\label{eq:v}
\end{equation}
which is $v=\sqrt2\,J$ at $J_r=\sqrt2$, as found numerically in
Sec.~\ref{sec:model}.

\paragraph*{Surface Green's function.}
Let $g(E)$ be the resolvent at the terminal site $D$ of a semi-infinite
critical chain that ends with a complete plaquette. Using Eq.~(\ref{eq:eff}),
$g$ obeys the two-step recursion of the effective alternating chain,
$g=[E-\epsilon-t_{\rm eff}^2/(E-\epsilon-t_x^2g)]^{-1}$, that is,
\begin{equation}
x\,t_x^2\,g^2-(x^2-t_{\rm eff}^2+t_x^2)\,g+x=0,\qquad x=E-\epsilon(E).
\label{eq:grec}
\end{equation}
Substituting Eq.~(\ref{eq:eff}) with $t_y=t_z$ and $t_x^2=2t_yt_z$, one
verifies by direct polynomial reduction that the solution of
Eq.~(\ref{eq:grec}) coincides identically in $E$ with the surface Green's
function of a uniform chain of hopping $t_x$,
\begin{equation}
g(E)=\frac{E-i\sqrt{4t_x^2-E^2}}{2t_x^2},\qquad |E|<2t_x .
\label{eq:gsurf}
\end{equation}
The lead that ends at a site $A$ has the same $g$, because the mirror
$A\leftrightarrow D$, $B\leftrightarrow E$, $C\leftrightarrow F$ maps $h$ to
$-h$ and the sublattice (chiral) symmetry maps $h$ to $-h$ as well; their
combination is a symmetry of $h$ that exchanges the two terminations. We
have checked Eq.~(\ref{eq:gsurf}) against the numerical decimation to
$10^{-9}$ over $|E|\le1$. For $t_y\ne t_z$ the identity holds only at $E=0$,
where $g(0)=-i/t_x$, which is all that the defect classification requires.

\paragraph*{Inter-plaquette bond.}
A single link of amplitude $\tau$ between two leads with terminal Green's
functions $g_L=g_R=g$ transmits
$T=4\tau^2({\rm Im}\,g)^2/|1-\tau^2g^2|^2$. With $\tau=\lambda t_x$ and
Eq.~(\ref{eq:gsurf}) this is
\begin{equation}
T_{\rm inter}(\lambda,E)=\frac{4\lambda^2s^2}{(1-\lambda^2)^2+4\lambda^2s^2},
\qquad s^2=1-\frac{E^2}{4t_x^2},
\label{eq:TinterE_app}
\end{equation}
the transmission of a single modified hopping in a uniform chain, which
reduces to Eq.~(\ref{eq:Tinter}) at $E\to0$.

\paragraph*{Intra-plaquette bonds.}
At $E\to0$ a bond factor $\lambda$ on one bond of one arm multiplies that
arm's contribution to $t_{\rm eff}(0)$ by $\lambda$ if the bond is $y$ or $z$
[the factor enters $h_{AB}$ or $h_{CD}$ in Eq.~(\ref{eq:elim})] and by
$1/\lambda$ if it is the middle $x$ bond [the factor enters $h_{BC}$ in both
numerator and denominator, and $E^2-\lambda^2t_x^2\to-\lambda^2t_x^2$ for
$E\ll\lambda t_x$]. The plaquette becomes a link of strength
$\mu\,t_x$ with $\mu=(1+\lambda)/2$ or $\mu=(1+\lambda)/2\lambda$,
respectively, in an otherwise uniform chain, and Eq.~(\ref{eq:TinterE_app})
at $E\to0$ with $\lambda\to\mu$ gives Eqs.~(\ref{eq:Tintra}) and
(\ref{eq:Tintrax}). For the $x$ bond the on-site terms
$\Sigma_{AA},\Sigma_{DD}\propto E/(E^2-\lambda^2t_x^2)$ of the weakened arm
are negligible only for $E\ll\lambda t_x$, which sets the range of validity
quoted in the text; the lattice values in Fig.~\ref{fig:transmissao}(a) were
computed at $E=10^{-4}$ and agree with Eq.~(\ref{eq:Tintrax}) down to
$\lambda=0.05$.

\paragraph*{Vacancy.}
Removing $C$ (the case of $B$, $E$, or $F$ follows by the mirrors) leaves $B$
attached to $A$ by the single bond $t_d=t_y$, and the arm $A$--$F$--$E$--$D$
as the only path through the plaquette, of strength $t_yt_z/t_x$ at low
energy. The dangling site contributes $\Sigma_{AA}=t_d^2/E$, which diverges
as $E\to0$. Treating the two sites $A$ and $D$ as the scattering region
between leads with $g=-i/t_x$, the Fisher--Lee transmission is
$T=4t_x^2|G_{AD}|^2$ with
$G_{AD}=t_{\rm arm}/[(E-t_d^2/E+it_x)(E+it_x)-t_{\rm arm}^2]$, and to
leading order in $E$
\begin{equation}
T(E)=\Big(\frac{2E\,t_yt_z}{t_x\,t_d^2}\Big)^2
=\frac{4E^2}{t_x^2}=\frac{E^2}{2},
\label{eq:Tvac_app}
\end{equation}
where the second equality uses $t_y=t_z=t_d$ and $t_x^2=2t_yt_z$, and the
third $t_x=2\sqrt2$. Equation~(\ref{eq:Tvac_app}) is the exponent-two law of
Eq.~(\ref{eq:Tvac}) with the amplitude $\tfrac12$ observed in
Fig.~\ref{fig:vacancia}(a).

\bibliography{refs}

\end{document}